\documentclass[twocolumn,superscriptaddress,aps,preprintnumbers,amsmath,amssymb,footinbib,prd]{revtex4}
\usepackage{graphicx}
\usepackage{graphics}
\usepackage{amssymb}
\usepackage{amsthm}
\def\bea{\begin{eqnarray}}
\def\eea{\end{eqnarray}}
\def\bec{\begin{center}}
\def\eec{\end{center}}
\newcommand{\beq}{\begin{equation}}
\newcommand{\eeq}{\end{equation}}
\newcommand{\nn}{\nonumber}
\newcommand{\tr}{\textrm{tr}}

\begin{document}
\title{Topological Dark Matter in the Little Higgs Models}
\author{Anosh Joseph}
\affiliation{Department of Physics, Syracuse University, Syracuse, NY 13244-1130 USA}
\author{S. G. Rajeev}
\affiliation{Department of Physics and Astronomy and Department of Mathematics, University of Rochester, Rochester, NY 14627-0171, USA}

\begin{abstract}
We show that certain little Higgs models with symmetry breaking $SU(N) \rightarrow SO(N)$ for $N \geq 4$ admit topologically stable solitons that may contribute to cosmological dark matter. We have constructed a spherically symmetric soliton and estimated its mass in the case of $SU(5) \rightarrow SO(5)$. Its lower bound is found to be around 10.3 TeV. Whether this particle is a fermion or a boson depends on the value of an integer-valued parameter of the underlying  theory, analogous to the number of colors of QCD. In either case, the particle is neutral. If it is a fermion, it is a Majorana particle, which could take part in a seesaw mechanism for neutrino masses.
\end{abstract}

\date{\today}
\maketitle
\preprint{SU-4252-893}

The hot Big Bang model of cosmology tells us that the Universe cooled down from a primordial hot and dense state to the present state of galaxies and other large-scale structures with a mean temperature of about 2.73 K \cite{Weinberg:2008zzc}. Certain theories of grand unification predict that the Universe underwent through a series of phase transitions as it cooled down, similar to what we observe in condensed matter systems. Phase transitions in the early Universe can give rise to certain stable configurations of matter known as topological defects. Different types of topological defects can arise depending on the symmetry breaking mechanism of the underlying field theory \cite{Vilenkin-Shellard}. They can appear in the forms of {\it magnetic monopoles}, {\it cosmic strings}, {\it domain walls}, {\it textures} and {\it skyrmions} \cite{Skyrme:1962vh}.

A field theory described by a continuous symmetry group $G$, when spontaneously breaks down to a subgroup $H \subset G$, the space of all accessible vacua of the theory called the {\it vacuum manifold} is defined to be the space of cosets of $H$ in $G$. The theory possesses a topological defect if some homotopy group of the coset space $\pi_d({\cal M} \equiv G/H)$ is nontrivial. When $d=0, 1, 2$ the defects respectively are domain walls, strings (vortices) and magnetic monopoles or textures. The case $d=3$, which plays a major role in this paper, gives rise to point-like topological defects called skyrmions.

Recently there has been much interst in a class of field theoretic models called the {\it Little Higgs} models \cite{ArkaniHamed:2001nc, ArkaniHamed:2002qy, ArkaniHamed:2002qx} in the context of weak scale symmetry breaking. These models provide a new logical possibility for natural electroweak symmetry breaking and a new partial resolution of the hierarchy problem in elementary particle physics. Introduction of new symmetries at the TeV scale by these models provides the cancellation of all quadratically divergent contributions to the Higgs mass at the one-loop level and pushes up the hierarchy problem to an energy scale of around 10 TeV. Little Higgs models have generated a lot of interest since any potential candidate to solve the hierarchy problem deserves serious attention.

Among the many possible ways of implementing the little Higgs paradigm, the {\it littlest Higgs} model \cite{ArkaniHamed:2002qy} is the simplest and most economical. This theory introduces a weakly coupled new physics at TeV energies, stabilizes the electroweak scale with a naturally light Higgs and is the smallest extension of standard model to date.

In this paper we address the interesting new possibility of bridging the natural electroweak symmetry breaking and cosmological dark matter - the non-baryonic, non-relativistic and weakly interacting matter that constitutes about 22\% of matter in the Universe. Since the Higgs particles appear as pseudo Nambu-Goldstone bosons in little Higgs models, skyrmion solutions that are stable and electrically neutral can also come out quite generically. We demonstrate the existance of a point-like, electrically neutral and topologically stable structure; a particle with a $Z_2$ charge; which could be a viable dark matter candidate. Its mass, estimated in the context of littlest Higgs model with T parity \cite{Cheng:2003ju}, is found to have a lower bound of around 10.3 TeV which is well below the unitarity limit \cite{Griest:1989wd} of viable dark matter particles. (Existence of other topological defects in the little Higgs model was investigated in \cite{Trodden:2004ea}.)

We start with a class of non-renormalizable effective field theories for pseudo Nambu-Goldstone bosons, in which a symmetry group $SU(N)$ is broken down to its real subgroup $SO(N)$ for $N \geq 4$. The case $N = 5$ is of most interest, as it appears in the little Higgs models.

At small energy scale compared to the symmetry breaking scale, the effective action has the form
\beq
\label{Action}
S_{1} = \frac{f^{2}}{8} \int d^{4}x \tr~\partial_{\mu}\Phi \partial^{\mu}\Phi^{\dagger} + \cdots,
\eeq
where $f$ is a parameter with dimension of energy and $\Phi$ is a scalar field given by a differentiable map from the Minkowsky space $R^{1, 3}$ to a nonlinear target manifold ${\cal M}_N$,
\beq
\Phi : R^{1, 3} \to {\cal M}_N.
\eeq
The target manifold ${\cal M}_N$ is the subset of symmetric matrices in $SU(N)$
\beq
\label{eq:targetspace}
{\cal M}_{N} = \{\Phi | \Phi = \Phi^{T},~\Phi \Phi^{\dag} = 1,~\det\Phi = 1 \}.
\eeq
It has a global symmetry $\Phi \to g\Phi g^{T},~g \in SU(N)$. Any $\Phi \in {\cal M}_{N}$ can be reduced to the identity by this transformation \cite{Schur}. That is, there is a $g \in SU(N)$ such that $\Phi=gg^{T}$. If we change $g \mapsto gh$, with $h \in SO(N)$, the product $gg^{T}$ is unchanged. Thus we can identify ${\cal M}_{N} = SU(N)/SO(N)$. The canonical projection to the cosets is $p : SU(N) \to {\cal M}_{N},~p(g) = gg^{T}$.

At spatial infinity, the field $\Phi(x)$ must approach a constant; the choice of this constant among all matrices satisfying Eq.(\ref{eq:targetspace}) will break the symmetry $SU(N)$ down to its real subgroup $SO(N)$. The parameter $f$ sets the scale of the symmetry breaking; in the little Higgs models it is expected to be a of the order of a TeV. The dots in Eq.(\ref{Action}) indicate that we are ignoring higher derivative terms, which are expected to be unimportant in the limit of `low' energies; that is, energies of the order of $f$.

Among the higher derivative terms we can add a new term that does not change the hyperbolic nature of the field equations and is still second order in time. This is the ``Skyrme term'' \cite{Skyrme:1962vh} given by
\beq
S_{2} = \frac{1}{32e^{2}}\int d^{4}x \tr~[\partial^{\mu}\Phi, \partial^{\nu}\Phi^{\dag}] [\partial^{\mu}\Phi, \partial^{\nu}\Phi^{\dag}]^{\dagger}.
\eeq

The value of the dimensionless constant $e$ depends on the details of the renormalizable theory of which Eq.(\ref{Action}) is the effective action. We will see that in the presence of this term, the effective action supports a topological soliton, whose mass is proportional to $M = fI/e$, with $I$ given in Eq.(\ref{eq:I}).

The more familiar Skyrme model \cite{Balachandran:1982dw, Witten:1983tx} is for the spontaneous breakdown of the symmetry $SU(2)\times SU(2)$ to $SU(2)$. The Nambu-Goldstone bosons are then the $\pi$-mesons. The action of the Skyrme model is then
\bea
S &=& \frac{f_{\pi}^{2}}{2}\int d^{4}x \tr~\partial_{\mu}g\partial^{\mu}g^{\dag} \nonumber \\
&&+ \frac{1}{32e^{2}}\int d^{4}x \tr~[\partial^{\mu}g, \partial^{\nu}g^{\dag}][\partial^{\mu}g, \partial^{\nu}g^{\dag}]^{\dagger} + \cdots
\eea
Closer in spirit to these papers are the Hopf soliton \cite{Faddeev:1996zj} (the case of ${\cal M}_2$) and even more so, the model studied in \cite{Balachandran:1982ty}, which is the case of ${\cal M}_3$.
\section{Topological Conserved Charge}
A continuous function $\Phi: R^{3} \to {\cal M}_N$ that approaches a constant at infinity can also be thought of as a map $\Phi : S^{3} \to {\cal M}_N$ by identifying the points at infinity. The homotopy group $\pi_{3}({\cal M}_N)$ has as elements the equivalence classes of such maps that can be deformed continuously into each other. It is well-known that \cite{Bryan:1993hz}
\beq
\pi_{3}({\cal M}_{2})=Z, \pi_{3}({\cal M}_{3})=Z_{4}, \pi_{3}({\cal M}_N)=Z_2, N \geq 4.
\eeq
The case $N = 3$ was studied in a different context some years ago \cite{Balachandran:1982ty}. We will focus here on the case $N \geq 4$, which includes the little Higgs models. There is just one non-trivial  equivalence class of maps $\Phi : R^{3} \to {\cal M}_N$; we will need to determine which representative of this class has the least energy. The non-trivial element of $\pi_3\left(SO(N)/SO(N)\right)$ is just the projection of the generator of $\pi_3\left(SU(N)\right)$.
\section{A Spherically Symmetric Ansatz}
Recall Skyrme's spherically symmetric (``hedgehog'') ansatz for a soliton of winding number one:
\beq
g_{2}(x)=e^{i\boldsymbol{\sigma}\cdot\boldsymbol{\hat{x}}\omega(r)},~\omega(0)=-\pi,~\omega(\infty)=0.
\eeq

The boundary conditions on $\omega$ ensure that the limits at $r=0,\infty$ are direction independent:
\beq
g_{2}(\infty)=1_{2},\quad g_{2}(0)=-1_{2}.
\eeq

This ansatz is spherically symmetric in the sense that a rotation in space can be compensated by the adjoint action of $SU(2)$:
\beq
g_{2}(R(A)x)=Ag_{2}(x)A^{\dagger},
\eeq
where $R : SU(2)\to SO(3)$ is the usual homomorphism. The obvious topologically non-trivial map $g_{2}g_{2}^{T}$ is not spherically symmetric: it is just cylindrically symmetric around the $x_{2}$-axis. This is because the representative $A$ of the rotation matrix does not cancel (is not orthogonal)  unless the rotation is around the $x_2$ axis. In fact, the energy minimizing configuration in ${\cal M}_2$ is only cylindrically symmetric \cite{Faddeev:1996zj}. If there is a spherically symmetric configuration, it is likely to have less energy.

There \cite{Balachandran:1982ty} is another spherically symmetric map $g_{3}: R^{3} \to SU(3)$, which interpolates between the identity at infinity and a cube root of unity at the origin:
\beq
g_{3}(\infty)=1_{3},~g_{3}(0)=e^{\frac{2\pi i}{3}}1_{3}.
\eeq
It is a generator or $\pi_{3}(SU(3$)). To construct it we start with the spherically symmetric ansatz
\bea
[g_{3}(x)]_{kl} &=& A(r)[\delta_{kl}-\hat{x}_{k}\hat{x}_{l}]+B(r)\epsilon_{kln}\hat{x}_{n}\nn \\
&&+C(r)\hat{x}_{k}\hat{x}_{l},~~\hat{x}_i \equiv x_i/|x|,
\eea
with the constraints $|C|=1, A^{*}B = B^{*}A, |A|^{2}+|B|^{2}=1$ to be unitary and $C(A^{2}+B^{2}) = 1$ to have determinant one. Under the action $g_3(x)\to Rg_3(Rx)R^T$ this is spherically symmetric.

So we get
\bea
A(r) &=& e^{-\frac{i}{2}\chi(r)}\cos\alpha(r),~C(r) = e^{i\chi(r)},\nn \\
B(r) &=& e^{-\frac{i}{2}\chi(r)}\sin\alpha(r).
\eea

The boundary conditions
\beq
\label{eq:boundary-c}
\chi(\infty)=0,~\chi(0)=2\pi/3,~\alpha(\infty)=0,~\alpha(0)=\pi
\eeq
ensure that the winding number is one. Computing $\Phi_{3}(x) = g_{3}(x)g_{3}^{T}(x)$,
\beq
[\Phi_{3}(x)]_{kl}=e^{-i\chi(r)}\delta_{kl}+[e^{2i\chi(r)}-e^{-i\chi(r)}]\hat{x_{k}}\hat{x}_{l}.
\eeq

Finally, we can embed  in  ${\cal M}_N$ to get a spherically symmetric representative for the generator of $\pi_{3}({\cal M}_N)$ for $N \geq 4$:
\beq
\label{eq:sph-symm}
\Phi_{N}(x)=\left(\begin{array}{cc}
\Phi_{3}(x) & 0\\
0 & 1_{N-3}\end{array}\right).
\eeq
For $\Phi_{N}(x)=g_{N}(x)g_{N}^{T}(x)$ with
\beq
g_{N}(x)=\left(\begin{array}{cc}
g_{3}(x) & 0\\
0 & 1_{N-3}\end{array}\right)
\eeq
and $g_{N}:R^{3} \to SU(N)$ has winding number one by the above construction. The configuration is spherically symmetric under the action $\Phi(x)\mapsto R\Phi(Rx)R^T$.
\section{Minimum Energy Soliton}
The mass of the solition in the theory with action $S_{1} + S_{2}$ will be the minimum of the energy
\beq
\label{eq:min-energy}
H(\Phi) \equiv f^{2}I_{1}(\Phi) + \frac{1}{e^{2}}I_{2}(\Phi),
\eeq
where
\bea
I_{1}(\Phi) &=& \frac{1}{8}\int d^3 x\tr~\partial_{i}\Phi \partial_i\Phi^{\dag},\\
I_{2}(\Phi) &=&\frac{1}{32}\int d^3 x\tr~[\partial_{i}\Phi, \partial_{j} \Phi^{\dagger}] [\partial_{i}\Phi, \partial_{j} \Phi^{\dagger}]^{\dagger},
\eea
among all functions $\Phi : R^{3} \to {\cal M}_N$ equivalent to the non-trivial element of $\pi_{3}({\cal M}_N)$. Since this topological charge is valued in $Z_{2}$, the topological soliton and its anti-particle are identical. In the absence of other interactions a single such soliton will be stable. Their number is not conserved - a pair of them can annihilate when they come in contact with each other.

As with skyrmions in QCD, it is clear that under a scaling $\Phi_{\lambda}(x)=\Phi(\lambda x)$, the two terms in the energy scale opposite to each other:
\beq
I_{1}(\Phi_{\lambda})=\lambda^{^{-1}}I_{1}(\Phi),~I_{2}(\Phi_{\lambda})=\lambda I_{2}(\Phi).
\eeq
Minimizing in the scale parameter, we see that the minimum energy will be proportional to $f/e$:
\beq
H_{min}=(f/e)\sqrt{I_{1}(\Phi)I_{2}(\Phi)}.
\eeq
We can make a variational estimate for the constant
\beq
\label{eq:I}
I=\min_{\Phi}\sqrt{I_{1}(\Phi)I_{2}(\Phi)}.
\eeq
The minimizing configuration should be invariant under the simultaneous rotation of the coordinate $x$ and a rotation by some $SU(2)$ subgroup (analogous to isospin in the Skyrme model) of $SU(N)$.

It is not difficult to make an estimate for the soliton mass $M$. Substituting the spherically symmetric ansatz Eq.(\ref{eq:sph-symm}) in Eq.(\ref{eq:min-energy}) and after some calculation we find
\bea
\label{eq:e_chi}
E (\chi)&=&\pi \int f^2 \Big[3r^2 \chi^{'2} + 4 (1-\cos 3\chi)\Big] dr\nn \\
&&+ \pi \int \frac{2}{e^2r^2}(1 - \cos 3 \chi)^2 dr\nn \\
&&+\pi \int \frac{2\chi^{'2}}{e^2}(3 - \cos 3 \chi -2\cos 6\chi)dr
\eea

We can find the minimum of energy in two ways, (i) by taking a variational ansatz for $\chi(r)$ or (ii) by solving $E(\chi)$ numerically. The variational ansatz gives an answer almost as good as the numerical solution.

\begin{figure}
\includegraphics[width=6cm]{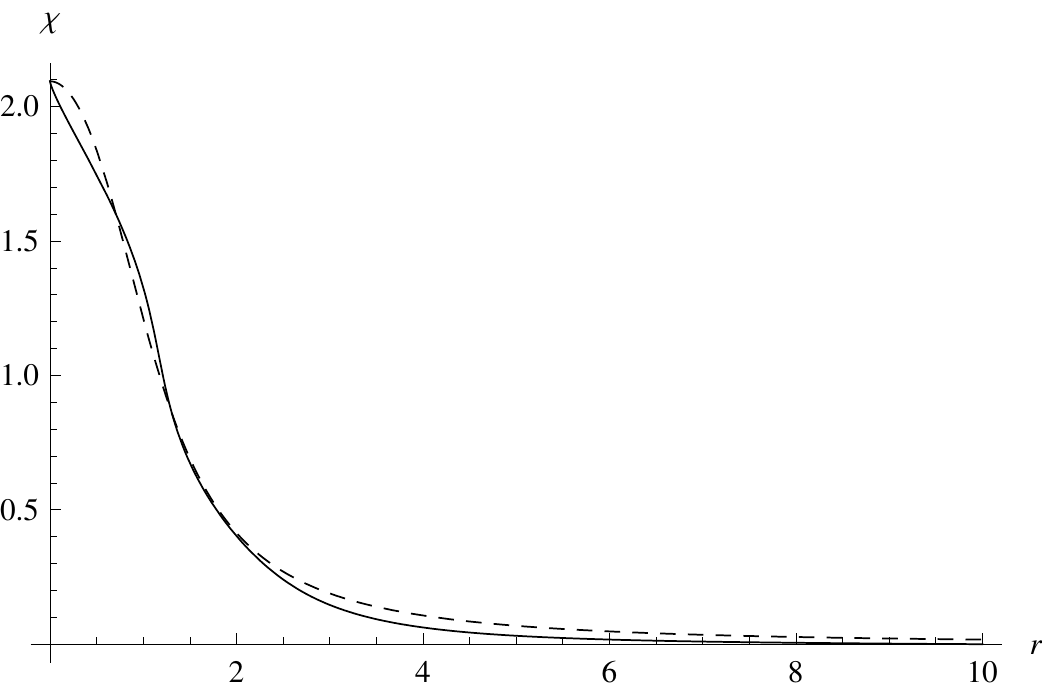}
\caption{The solution for $\chi(r)$ in units where $e=f=1$.}
\label{chi-plot}
\end{figure}

We tried the following ``stereographic" ansatz for $\chi(r)$
\beq
\chi(r) = \frac{4 \pi}{3} \arctan\Big(\frac{R^n}{r^n}\Big).
\eeq
for $n=1,2,3,4$. They satisfy  the boundary conditions given in Eq.(\ref{eq:boundary-c}). The lowest value for energy was obtained for $n=2$ and
$
R = R_0 \approx \frac{1.13}{ef}.
$
The value of the minimum energy is
$
E(R_0)\equiv M = 105 \frac{f}{e}.
$
The numerical solution of the differential equation for $\chi$ gives a slightly lower value of energy close to the variational ansatz: 
\beq
E = 102.8 \frac{f}{e}.
\eeq
We plot the solution in Fig. \ref{chi-plot} in units where  $e=f=1$; the dashed curve is the variational ansatz with $n=2$ and the solid curve is the numerical solution.
We need an estimate for the value of the dimensionless constant $e$ as well as $f$ to get a number for the mass of the soliton $M$.  Precision electroweak constraints put a lower bound on the symmetry breaking scale $f$ typically of about 500 GeV \cite{Hubisz:2005tx} for little Higgs models with T parity \cite{Cheng:2003ju}. The Skyrme constant is in principle determined by the underlying renormalizable theory of which the little Higgs model is the effective theory. At the moment we do not know what this effective theory is; even if we knew it, we do not yet know how to compute such constants. But it is reasonable to expect that $e$ will have the same order of magnitude as for QCD; this is the best we can do with our current knowledge. In QCD $e\approx 5$, as we can deduce from the value of the nucleon mass. With these values we get  an estimate (a lower bound for $e=5$) for  the mass of the soliton
\beq
M \gtrsim 10.3~\textrm{TeV}.
\eeq

Since the mass of this particle is below the unitarity bound ($\lesssim$ 340 TeV) \cite{Griest:1989wd}, it cannot be excluded from the list of viable dark matter candidates. Possible cosmological implications such as relic abundance, decay \footnote{These particles can be meta-stable with very long life times since the global symmetry in the little Higgs models is approximate.} and annihilation cross-sections of these particles should be explored.

The coefficient $N_c$ of the Wess-Zumino-Witten term (which is equal to the number of colors of QCD) determines whether the baryon is a boson or a fermion: for odd $N_c$ it is a fermion and for even $N_c$ it is a boson. We do not yet know if  the  analogous parameter in the little Higgs models is even or odd: both possibilities  would give the same effective theory at the electroweak scale. When our skyrmion  is a boson,  it can be represented by a real scalar field $S$ whose couplings have the discrete symmetry $S\to -S$. The conserved quantity associated to this symmetry is the topological charge of the little Higgs model. The phenemenological consequences of such scalars have been investigated in \cite{Davoudiasl:2004be}.  When our particle  is a fermion, it is a Majorana particle.  In this case it could be the fermion responsible for the neutrino masses in a seesaw mechanism \cite{Mohapatra:1980yp}. To flesh out this idea, we need to understand the mixing matrix of the topolgical soliton with neutrinos, induced by the anomalous  coupling of neutrinos of the bosons of the little Higgs models \cite{Kaymakcalan:1983qq, Hill:2007nz}.
\begin{center}
{\bf Note added}
\end{center}
While this manuscript was in preparation a preprint with a significant overlap with the work presented here appeared in the arXiv \cite{Murayama:2009nj}.

\section{acknowledgments}
A. J. thanks Jay Hubisz for useful discussions and the Department of Physics and Astronomy, University of Rochester for their friendly hospitality. S. G. R. thanks Fred Cohen for discussions  on homotopy groups, Mark Trodden on cosmology, and A. P. Balachandran on solitons. We also thank Marc Gillioz for pointing out the numerical error in Eq (\ref{eq:e_chi}). S. G. R.'s work is supported in part by a grant from the U. S. Department of Energy under the contract number DE-FG02-91ER40685. A. J.'s work is supported in part by the U. S. Department of Energy grant under the contract number DE-FG02-85ER40231.

\end{document}